# Progress of microscopic thermoelectric effects studied by micro-and nano-thermometric techniques

Xue Gong[1], Ruijie Qian[1], Huanyi Xue[1], Weikang Lu[1], Zhenghua An[1,2,3]*

[1] *State Key Laboratory of Surface Physics, Department of Physics, Fudan University, Shanghai 200433, China*

[2] *Institute for Nanoelectronic Devices and Quantum Computing, Fudan University, Shanghai 200433, China*

[3] *Shanghai Qi Zhi Institute*，*41th Floor, AI Tower, No. 701 Yunjin Road, Xuhui District, Shanghai, 200232, China*

*Corresponding author. E-mail: \*anzhenghua@fudan.edu.cn*

Heat dissipation is one of the most serious problems in modern integrated electronics with the continuously decreasing devices size. Large portion of the consumed power is inevitably dissipated in the form of waste heat which not only restricts the device energy-efficiency performance itself, but also leads to severe environment problems and energy crisis. Thermoelectric Seebeck effect is a green energy-recycling method, while thermoelectric Peltier effect can be employed for heat management by actively cooling overheated devices, where passive cooling by heat conduction is not sufficiently enough. However, the technological applications of thermoelectricity are limited so far by their very low conversion efficiencies and lack of deep understanding of thermoelectricity in microscopic levels. Probing and managing the thermoelectricity is therefore fundamentally important particularly in nanoscale. In this short review, we will first briefly introduce the microscopic techniques for studying nanoscale thermoelectricity, focusing mainly on scanning thermal microscopy (SThM). SThM is a powerful tool for mapping the lattice heat with nanometer spatial resolution and hence detecting the nanoscale thermal transport and dissipation processes. Then we will review recent experiments utilizing these techniques to investigate thermoelectricity in various nanomaterial systems including both (two-material) heterojunctions and (single-material) homojunctions with tailored Seebeck coefficients, and also spin Seebeck and Peltier effects in magnetic materials. Next, we will provide a perspective on the promising applications of our recently developed Scanning Noise Microscope (SNoiM) for directly probing the non-equilibrium transporting hot charges (instead of lattice heat) in thermoelectric devices. SNoiM together with SThM are expected to be able to provide more complete and comprehensive understanding to the microscopic mechanisms in thermoelectrics. Finally, we make a conclusion and outlook on the future development of microscopic studies in thermoelectrics.

## Contents

# 1 Introduction

Moore's law predicts that the development direction of electronic industry tends to be more miniaturized and integrated to realize extraordinary functions. Then the arising problems, such as quantum tunneling effect and locally overheated hotspots, become more severe. While most of works focus on the quantum effect, the heat problem turns out to be also a bottleneck for further development and can easily destroy the device performance, reliability and stability [1-5]. What's more, the large amount of power consumed by integrated circuits exacerbates the energy crisis and deteriorates environmental problems. Typically heat in solid-state electronics can be dissipated by passive and/or active cooling [6-11], or potentially reused through thermoelectric conversion which remains although extremely challenging so far. Passive cooling refers to the use of the intrinsic thermal conductivity property of materials for spontaneous heat dissipation, and active cooling refers to the use of external refrigeration devices for active heat exchange, pumping heat from the high heat generation regions to the cold surrounding environment or reservoir. As the size of the device decreases and the integration density creases, conventional passive cooling for the high-density of heat generation in integrated chips becomes not sufficient any more, if relying on the finite thermal conductivities of available natural materials. Possible efforts may include utilizing new highly thermal conductive materials, or developing artificial thermal metamaterials [12-14], or introducing additional radiative cooling methods [15, 16]. These passive approaches are, however, not yet applicable for real integrated devices and also the possible improvements are expected to be practically moderate. Therefore, active cooling through thermoelectric Peltier effect appears to be a very competitive and cooperative solution. Figure 1 depicts a conceptual scheme utilizing both passive and active cooling. On the one hand, the heat-generating device is mechanically supported by the heat-conducting substrates which can be engineered into thermal metamaterial and therefore managed to control the heat flow; on the other hand, the heat-generating device can be designed to be in contact with the cold side of a Peltier cooler such that the heat produced inside the device can be pumped actively to the hot side and then dissipated with passive or fan cooling techniques. To be more efficient and cost-effective, the cold-side of Peltier cooler and its thermal contact may be optimized to match the spatial profiles of the hotspot distribution in heat-generating device. Noting that thermoelectric effects are a smart approach with the advantages of environmentally friendly, no mechanical moving parts, no vibrational noise, relatively small volume and potentially compatible with integrated nanofabrication technology, this combined approach therefore provides a promising and eventual solution to optimize the cooling [17]. It will be best if the dissipated energy can be subsequently harvested and thermoelectrically reused. The above scenario, however, remains hardly realistic for integrated devices before the thermoelectric efficiencies are significantly improved and thermoelectric device are considerably downsized. To this aim, visualizing and understanding the microscopic mechanisms in thermoelectricity particularly in nanoscale is of great importance for actual application in nanoscale heat management. Besides, the scientific significance of nanoscale thermoelectricity research arises from the fact that, when the characteristic length of the object is reduced to be comparable to or even smaller than the mean free path (MFP) of the energy carriers ( phonons or electrons), extraordinary properties such as quantized thermal conductance [18], ballistic phonon transport [19],suppressed thermal conductance [20, 21], decoupled electron and heat [22], will be exhibited, which are yet to be directly visualized and further explored in real space.

The technological challenge in studying nanoscale thermoelectricity lies in the extreme difficulties in directly probing nanoscale heat in solid-state devices [23-26]. Same as in macroscopic scale, the nano-heat can be detected either based on the thermodynamic law by direct contact measurement or radiatively in a noncontact manner. The former contact method requires precise fabrication of local nanothermometer as well as associated read-out electronics and the latter noncontact one has to overcome the diffraction limit in order to realize nanometer spatial resolutions.

In this short review, we concentrate on recent progress in studying the micro-/nano-scale thermoelectric effects in various nanomaterial systems and the content is organized as follows. First, in Section 2, we briefly introduce the thermometric techniques and focus on the operating mechanism of scanning thermal microscope (SThM) which facilitates experimental studies of nanoscale thermoelectric effects. Second, in Section 3, we highlight some recent significant applications of SThM in investigating microscopic thermoelectric effects in both two-material heterojunctions and single-material homojunctions, respectively. The origin of the thermoelectric effects in a single material is stressed to be the modification of the local Seebeck coefficient, paying attention to how to image the surface Seebeck coefficient distribution. Newly reported spin Seebeck and Peltier effects in magnetic materials are also presented which have triggered the rapid development of spin caloritronics. In Section 4, we discuss the study of probing non-equilibrium transporting hot charges (instead of lattice heat) with our recently developed Scanning Noise Microscope (SNoiM) and give a perspective on the combination of SNoiM and SThM for indepth understanding of the microscopic mechanisms in thermoelectrics. Finally, we make a short conclusion and outlook for the future direction in this research field.

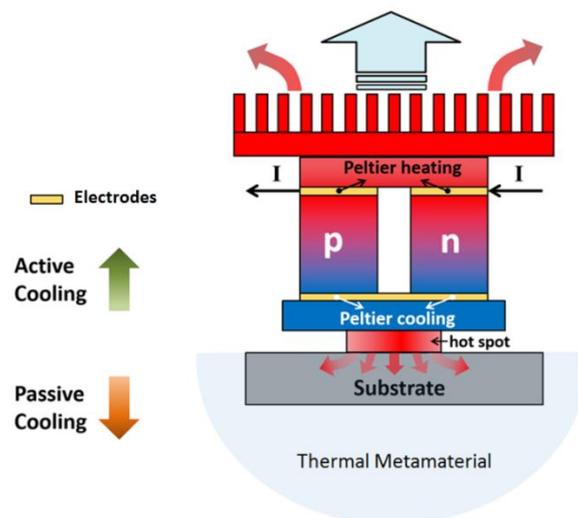

**Fig. 1** Conceptual diagram of employing both active and passive cooling for solid-state heat generating devices

## 2 Working principle of SThM

SThM is one of the important modules based on atomic force microscope (AFM) with temperature sensing capability, which was first invented by Williams and Wickramasinghe [27] in 1986 for the purpose of imaging the surface topography of an insulator by the using the thermal signal as feedback. The general configuration for SThM is integrating a temperature sensor (i.e., nanothermometer) at the tip of the probe and the typical temperature sensors adopted are thermistors, thermocouples, Schottky diodes and even superconducting quantum interference devices [28-31]. Among them, thermocouple can be rather easily fabricated and perform reliably with good sensitivity and is therefore widely used in many commercial setups just as it is often used in our research. By scanning the nanothermometer on top of the target sample, the two-dimensional temperature profiles can be recorded straightforwardly together with the sample topography. It is worth mentioning that the scanning Joule expansion microscopy (SJEM) [32, 33] discussed later is also one kind of SThM since it is also based on the probe scanning to detect local temperature although the scanned tip itself is not a nanothermometer.

The working principle of SThM is the same as that of AFM, except that the tip is functionalized thermometrically and correspondingly the electronic module has to be upgraded for thermal detection [34]. As schematically shown in Figure 2(a), when the tip is scanned along the sample surface, the laser focused onto the probe cantilever will be deflected due to the change of the interaction force between the tip and the sample surface to be measured. The deflected signal will be collected by the photodiode system as the feedback for AFM tip height control. The operating mode is generally contact mode, but tapping mode or peak-force mode are also possible and the operating environment can be either atmospheric or vacuum and/or cryogenic. Various experiments have proved that the signal under vacuum condition is better and cleaner than in air. Since the heat exchange between the tip and the sample surface is dominated by the direct contact, the spatial resolution of SThM is typically determined by the tip size. Experiments show that the resolution of SThM can reach about 10nm, which is more than enough for detecting typical thermal phenomena at the nanoscale.

The thermocouple temperature detection module (Figure 2(b)) used in SThM is a representative application of the thermoelectric Seebeck effect mentioned below (equation (1)). The feedback laser heats the cantilever to a stable reference temperature. When the tip gradually approaches to the sample surface and eventually touches it, the temperature difference between the sample surface and the tip turns on the heat flow through solid-solid contact and a temperature gradient related to the reference temperature will be established, producing a measurable electric voltage. The thermovoltage of each location on sample surface will be recorded after passing through a preamplifier and give the final temperature imaging, with the topography acquired simultaneously from another channel. Figure 2(c) exhibits the simultaneously measured approach curve and the corresponding thermal signal. When the probe is brought to contact with the sample surface, the thermal signal of the probe jumps from $T_{nc}$ to $T_c$ due to probe-sample contact heat conduction. To study the thermoelectric effects in current-carrying devices, lock-in technique can be utilized to overcome the background temperature fluctuation and therefore increase the temperature sensitivity. In this case, the device to be measured can be biased with alternating current (AC) voltage and then the thermovoltage can be demodulated at same frequency or high harmonic frequencies, depending on which physical effect is to be tested. This SThM with additional lock-in technique provides an agile tool to study the thermoelectric effects in electronic devices and, most importantly, with nanoscale spatial resolution.

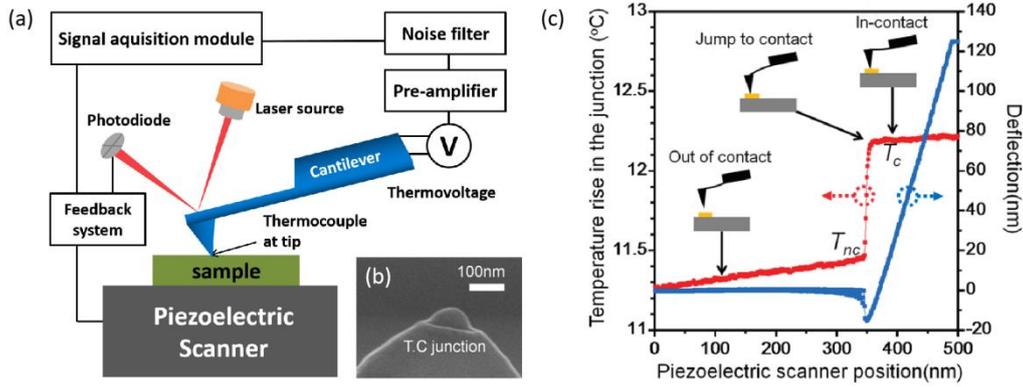

**Fig. 2** (a) Experimental setup of scanning thermal microscopy (SThM). (b) The scanning electron micrograph (SEM) of the Cr/Au thermocouple probe. (c) The approach curve of the probe closes to sample surface and the simultaneously measured thermal signal curve. When the probe is brought to contact with the sample surface, the thermal signal of the probe jumps from $T_{nc}$ to $T_c$ due to probe-sample contact heat conduction. Reproduced from Ref. [48].

In addition to SThM, there are also various other functional micro-/nano-scale temperature measurement techniques. Among them, optical method working in non-contact mode such as thermoreflectance thermal imaging microscopy and the infrared thermal microscope mentioned below have spatial resolution of ~100nm and ~5-10μm, respectively. Compared with the infrared thermal microscope which is limited by the optical diffraction limit, the thermoreflectance thermal imaging microscopy has a better spatial resolution due to the use of a shorter wavelength. Although the resolution is not as good as SThM, these two techniques possess high imaging speed and sensitivity, high temperature resolution. Despite the SThM has ultra-high resolution, its heat transfer mechanism is complicated and sophisticated calibration is needed for quantitative analysis. Besides, its scan speed is slow due to the fact that it takes time for the tip to reach thermal equilibrium with the sample under test. For the detailed description of these different techniques and their comparison, please refer to these existing reviews [23-26]. In the following, we will focus on the recent experimental progress in studying nano-scale thermoelectric effects acquired on different material systems.

## 3 Researches on micro-/nano-scale thermoelectric effects

Thermoelectric effects typically include two opposite effects, i.e., Peltier effect and Seebeck effect, referring to the reversible conversion between temperature difference and electric voltage [35]. Seebeck effect states that a voltage drop will be produced when a temperature difference is formed in a conductor and can be expressed as:

$$V = S * \nabla T \qquad (1)$$

where $V$ is thermovoltage, $S$ is the Seebeck coefficient or termed as thermopower which is generally related to the properties of the material itself and $T$ is the temperature with $\nabla T$ being its gradient. The Peltier effect is contrary to Seebeck effect and refers to voltage gradient (i.e., electrical current) generating a temperature difference. The representative applications of these two effects in macroscopic scale are thermocouples and refrigerators, respectively.

In this section, we review some typical researches of nanoscale thermoelectric effects observed in various material systems including two-dimensional materials (e.g., graphene), semiconductor nanowires (e.g., InAs nanowires), and phase change memory materials (e.g., $Ge_2Sb_2Te_5$), p-n homojunction structures, and novel single-material thermoelectric structures. A burgeoning field called spin caloritronics is also presented, focusing on the spin thermoelectric effects.

## 3.1 Heterojunction thermoelectric effects

In a closed circuit composed of two homogeneous materials, when there is a direct current (DC) energy transfer occurs, one contact will produce heat and the other contact will absorb the heat and therefore producing temperature difference between the two contacts because of the thermoelectric Peltier effect. This Peltier effect was discovered by French scientist Peltier in 1834. The physical explanation behind this is that the charge carriers are at different energy levels in different materials, when current flows through the node, in order to maintain energy and charge conservation, they must exchange energy with the environment. Energy is absorbed or released in the form of heat at the interface of the two materials and the heat flux can be written in the form of

$$\dot{Q} = (\Pi_A - \Pi_B) \times I \tag{2}$$

where $\Pi = T \times S$ is the Peltier coefficient, with the subscript corresponding to material A and B, respectively. The Peltier coefficient is linked to Seebeck coefficient $S$ through Thomson-Onsager relation. $I$ is the current flow through the conductor. It can be seen from the formula that whether heat is absorbed or released at the junction is determined by the direction of current flow and the sign of the Seebeck coefficient.

At present, most of thermoelectric applications are mainly based on conventional semiconductors such as PbTe, $Bi_2Te_3$ and their alloys, $Sb_2Te_3$, SiGe, InSb, skutterudite materials, half-Heusler alloys, and usually consist of several p- and n-type legs which are connected electrically in series to add-up all thermoelectric cooling capacities [36, 37]. In the field of emerging novel materials, in 2018, Jin *et al*. experimentally studied the Peltier effect at the interface between organic film and metal, taking advantage of an infrared thermal microscope with a spatial resolution of several microns [38]. The experimental and simulation results reveal that organic thermoelectric (OTE) materials have great application prospects in the future. With the emergence of nanomaterials and continuous reduction of device sizes in the last few decades, remarkable progress of micro- and nano-processing technologies has been made. Correspondingly, Hicks and Dresselhaus proposed that the use of quantum confinement effects can greatly improve thermoelectric efficiency by reducing the dimensions of materials, which stimulate broad interest in the study of micro-/nano-scale thermoelectric effects, and seeking new materials as well as novel mechanisms for boosting thermoelectric energy conversion and cooling efficiencies [35, 39].

Graphene as a promising two-dimensional material first found in 2004 has been or will be widely applied in electronic and photonic devices. However, the performance of graphene-based electronics is greatly compromised due to contact problems with metal electrodes [40, 41]. In 2011, Grosse *et al*. used an equipment called scanning Joule expansion microscopy (SJEM) (Figure 3(a)), based on the lattice thermal expansion arising from local electrothermal hotspot. This technique is applied to quantitatively study the respective proportions of Joule heating, current crowding, and thermoelectric effect at the interface between graphene and metallic electrodes (Figure 3(c)) [42]. Two years later, the same research group observed similar thermal phenomenon at the junction of the phase change memory material $Ge_2Sb_2Te_5$ (GST) and the metal TiW electrode by SJEM (Figure 3(b)) [43]. The above two experimental results clearly verify and quantify the existence and the influences of thermoelectric heating and cooling at the graphene/metal and GST/metal heterojunctions.

The SJEM technique measures the thermal expansion signal of the sample surface caused by self-heating due to external current excitation, and the expansion signal is a convolution of the surface topography height, the local expansion coefficient and the temperature rise. The above two experiments using SJEM both require to be spin-coated a poly (Polymethyl methacrylate) (PMMA) on the surface to eliminate the influence of sample morphology and local thermal expansion coefficient differences. The major advantage of leveraging SJEM is that it will not be affected by the complicated mechanism of heat transfer between the tip and the sample. And it does not require to reach a steady thermal equilibrium state, therefore allowing faster tip scanning speed and shorter imaging time.

As a comparison, in 2016, Vera-Marun *et al*. proposed an electrical measurement method to directly measure thermoelectric cooling and heating at the junction of graphene and Au electrode [44]. As shown in Figure 3(d), they made an integrated thermocouple, which is composed of NiCu with a large Seebeck coefficient and the Au electrodes with a small Seebeck coefficient, which also served as the source and drain of graphene channel. When applying an electrical current between the source and the drain, the NiCu/Au thermocouple around the junction of Au electrode and graphene will produce a thermovoltage change due to local temperature variation. The measured thermovoltage change can be converted into temperature through the Seebeck coefficient $S_{NiCu/Au}$ of the thermocouple calibrated beforehand, and the sensitivity of this experimental equipment can reach the order of mK. They used this setup to systematically study the effects of modulating carrier type and carrier density on the behavior of thermoelectric effects by adjusting the gate voltage (Figure 3(e), (f), (g)). Thanks to the temperature sensitivity of integrated thermocouple, the maximum temperature rise measured at room temperature was about 15mk in this work. Despite of the easy operation of this technique, this pure electrical measurement method lacks the two-dimensional imaging capability, possibly influenced by the device geometry, size and soon, and therefore cannot achieve the spatial resolution as required by most of desired experiments in this field.

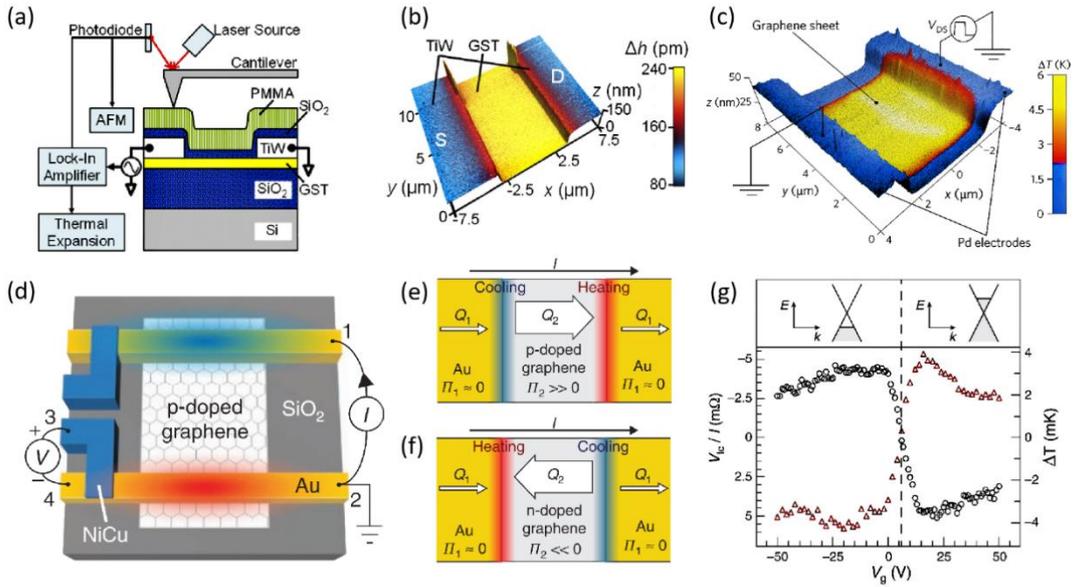

**Fig. 3** Experimental setup and operating principle of scanning Joule expansion microscopy (SJEM) for investigating the (b) GST device and (c) graphene-metal contacts. (d) Schematic diagram of directly electrical temperature measurement of graphene-Au contact with NiCu/Au thermocouple. (e) and (f) are the description of the Peltier effect at graphene-Au contacts corresponding to p-doping graphene and n-doping graphene through field effect regulation. (g) The dependence of the measured thermocouple signal (left axis) and the converted temperature (right axis) through the calibrated $S_{NiCu/Au}$ on the gate voltage. Reproduced from Refs. [42, 43, 44].

To combine the scanning imaging capability and the excellent temperature sensitivity of thermocouple to achieve better performance for nanoscale thermoelectric study, researchers have tried to implement SThM with thermocouple tip (somewhat a combination of the above SJEM and pure electrical method with thermocouple electrodes) in an ultrahigh vacuum environment, and also proposed double scanning technique, null point method, etc. [45-49]. One early work was reported by Menges *et al*. for directly measuring nanoscale thermoelectric effects with SThM [50]. The employed method can overcome the influence of tip-sample contact-related artifacts on the thermal signal, and only needs to simply calibrate the Seebeck coefficient of the probe used. This method was first applied to measure the thermoelectricity properties of the contact interface between InAs nanowire and Au electrodes. As illustrated in Figure 4(a), a bipolar sine wave with a certain amplitude $V_{PP}$=0.7V and frequency $f$=10kHz is input between the source and drain of the InAs nanowire, thus leading to the temperature oscillating at the source frequency $f$. Since Joule heating has a square dependence on the current, and Peltier heating and cooling have a linear dependence on the current, the extracted first ($f$) and second ($2f$) harmonic signals by the use of lock-in amplifier correspond to the Peltier and Joule signals, respectively. It demonstrates that this SThM approach allows us to study the two effects separately compared with the previous measurement method, which only measures the temperature field distribution mixed with all the effects. The measured temperature maps are displayed in right panels of Figure 4(b) (Peltier heating/cooling) and Figure 4(c) (Joule heating). It is apparent that the Peltier heating/cooling appear exactly at the interfaces between the nanowire and the electrodes, with zero signal at the middle position of the nanowire.

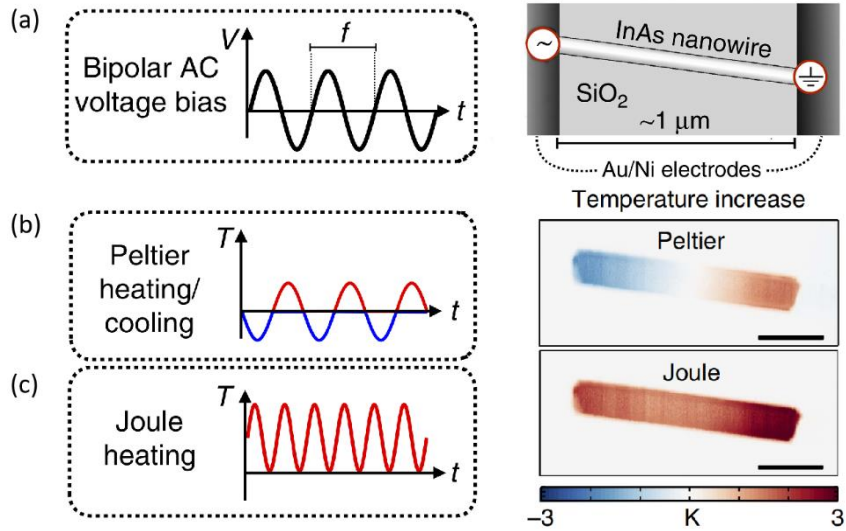

**Fig. 4** (a) Schematic diagram of the principle of measuring the thermoelectric effects at the interface between InAs nanowire and gold electrodes. A sine wave with a certain amplitude 0.7V and frequency $f$ is input to the sample, which results in the sample temperature oscillates at the frequency $f$. (b) The measured signals are extracted by lock in amplifier at two different frequencies (1$f$ and 2$f$) corresponding to Peltier (b) and Joule heating (c) effects, respectively. Reproduced from Ref. [50].

**3.2 Homojunction thermoelectric effects**

In addition to the microscopic thermoelectric effects found at the interface composed of two dissimilar materials, recent works have also observed similar phenomena on single material system which may find potential application in future due to its unique advantages such as simpler structure, no interface and low production cost. The existence of thermoelectric effect (in particular Peltier effect) in single material systems is somewhat surprising as the conventional thermoelectric effect requires a junction between two dissimilar materials. So the first subsection of this part will be dedicated to the discussion of the underlying mechanisms for the thermoelectric effect observed in a single material and to the methods of modifying Seebeck coefficient and the experimental techniques of local detection of Seebeck coefficient distribution. After that, we will focus on related researches mainly using SThM.

### 3.2.1 Seebeck coefficient modification

From the definition of thermoelectric effects, we can infer that as long as there is an interface consisting of materials with different Seebeck coefficients, the thermoelectric effects can take place. Therefore, the novel concept of a single-material thermocouple and the single material thermoelectric effects have emerged under the condition that the modification of local Seebeck coefficient becomes possible. The semi-classical Mott expression of the Seebeck coefficient can be written as:

$$S = -\frac{\pi^2 \kappa_B^2 T}{3e} \frac{d \ln \sigma(E)}{dE}\Big|_{E_F} \tag{3}$$

Here $e$ is the elementary charge, $\kappa_B$ is the Boltzmann constant, T is the temperature, $\sigma(E) = n(E)e\mu(E)$ is the energy-dependent conductivity taken at Fermi energy $E_F$ and $\sigma(E) = n(E)e\mu(E)$ is the carrier density that fill the energy states between $E$ and $E+dE$, $g(E)$ is the density of states of electrons in unit energy interval near energy E, $f(E)$ is the Fermi distribution function and $\mu(E)$ is the mobility. The equation suggests that the $S$ is very sensitive to the asymmetry of the density of states near the $E_F$ and is linearly dependent on temperature.

To modify Seebeck coefficient, nanostructuring and energy band engineering are two alternative strategies to locally engineer the Seebeck coefficient distribution [51, 52]. The quantum confinement of carriers in low-dimension materials can largely increase Seebeck coefficient which has been reported by several groups in various systems such as one-dimensional quantum wires, two-dimensional wells, superlattices nanocomposite materials [53]. By flexibly adjusting the width and thickness of the quantum well, the diameter and doping concentration of the nanowires, and the nanocrystal sizes of the superlattice, the Seebeck coefficient can be several times larger than that of bulk materials. The physical reason for the substantial increase in Seebeck coefficient is due to sharp features in the electronic density of states caused by quantum confinement. In addition, carrier energy filter is an effective mechanism that can be employed to improve Seebeck coefficient in low-dimension nanostructuring or nanocomposite materials. By artificially introducing high barriers in the conduction band or valence band for n-type or p-type semiconductors respectively, the local Seebeck coefficient can be changed. Same effect often occurs near the grain boundaries in nanocomposites, which can screen out high energy electrons that carry more energy to pass through, thus will produce a larger Seebeck coefficient [54-56].

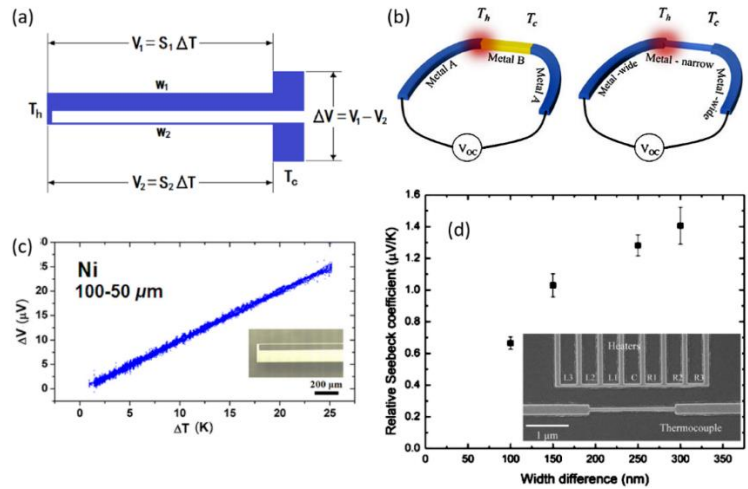

**Fig. 5** The concept and experimental setup of verification of the difference in Seebeck coefficient of a single material caused by the size effect proposed by Sun *et al.* (a) and Szakmany *et al.* (b), respectively. (c) and (d) are the corresponding experimental results, the inset in (c) and (d) are the corresponding optical image and SEM image of each device, respectively. Reproduced from Refs. [58, 59].

In addition to the quantum confinement approach, the classical size effect will also be able to adjust the Seebeck coefficient, which mainly originates from the change of mean free path (MFP) caused by geometrically confined scattering [57]. In 2011, Sun *et al.* found that the Seebeck coefficient of thin film stripe strongly depends on its width [58]. The designed structure is shown in Figure 5(a) and the optical picture of the actual sample is shown in the inset in Figure 5(c). Two strips of different widths are connected at one end and disconnected at another end. When there is a temperature gradient between the two ends, an open circuit voltage will be measured at the disconnected end. This experimental design idea eliminates tedious experimental preparation and testing steps while verifying the size dependence of Seebeck coefficient. It is worth noting that the width of stripes is much larger than the MFP. Hereafter, Szakmany *et al.* applied this concept and demonstrated single metal nanoscale thermocouple (Figure 5(b)), the experimental results (Figure 5(d)) show that the relative Seebeck coefficient decreases as the width difference decreases and which is in good agreement with the Fuchs-Sondheimer model [59, 60].

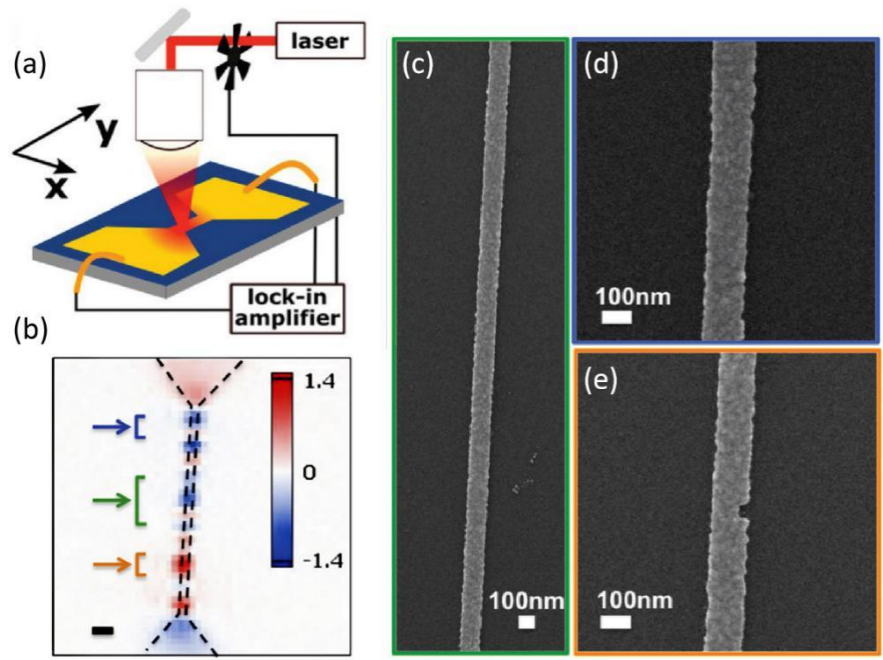

**Fig. 6** (a) The experimental setup of photothermoelectric (PET) measurement of a thin-film gold nanowire. (b) The measured PET image. (c), (d) and (e) are the scanning electron micrographs (SEM) corresponding to the different colored markers in (b). Reproduced from Ref. [62].

Subsequently, Pavlo Zolotavin conducted a detailed study to extrapolate the photothermoelectric (PET) properties of thin-film gold nanowires using scanning laser microscopy (Figure 6(a)). The focused beam could generate local hot spot due to the direct light absorption of the metal nanowires via plasmon resonance. In long nanowires, inhomogeneous spatial profile of the PET signal has been observed (Figure 6(b)) which is mainly ascribed to the change of the local Seebeck coefficient induced by the width variation, structural defects, grain structure and surface conditions which are clearly seen by the SEM images exhibited in Figure 6(c), (d) and (e) corresponding to different area marked in Figure 6(b) [61-63].

Another interesting finding is that the field-effect gating is a powerful degree of freedom in low-dimensional systems not only for tuning the Fermi level and therefore carrier density, but also can effectively engineer the energy band structure without introducing a great number of disorders or structural changes, thereby manipulating the Seebeck coefficient [64-66]. For example, in 2009, Liang *et al.* first demonstrated that the Seebeck coefficient of a single PbSe nanowire can be modulated by the electric field-effect and the room temperature value of Seebeck coefficient can be increased by 3 times continuously from 64 to 193 $\mu$V/K with sweeping the gate voltage [67]. In an n-type semiconductor, the field-effect gating can effectively and precisely tune the electron density in the conduction band, thereby changing the Fermi level of the nanowire and hence changing the Seebeck coefficient in the way that the above Mott formula (equation (3)) expects. Subsequently in 2016, Shimizu *et al*. and Saito *et al*. successively controlled the Seebeck coefficient in metallic and semiconducting single-walled carbon nanotubes and black phosphorus (BP) single crystal sheets (~40nm) by the use of ionic gating. Figure 7(a) and Figure 7(b) are the corresponding experimental diagram and optical image of BP single crystal sheets, respectively [68, 69]. The Seebeck coefficient value of BP single crystal sheets can reach 510 $\mu$V/K at 210K (Figure 7(c)) and in the original paper they also pointed out that by reducing the thickness of black phosphorus, its Seebeck coefficient can be further increased. From these existing studies including theories and experiments, it can be seen that gating plays a great role in promoting the application of low-dimensional materials in thermoelectricity and the above three materials with various advantages are theoretically expected to be excellent candidates for thermoelectric applications despite that more experimental investigations are still needed [70].

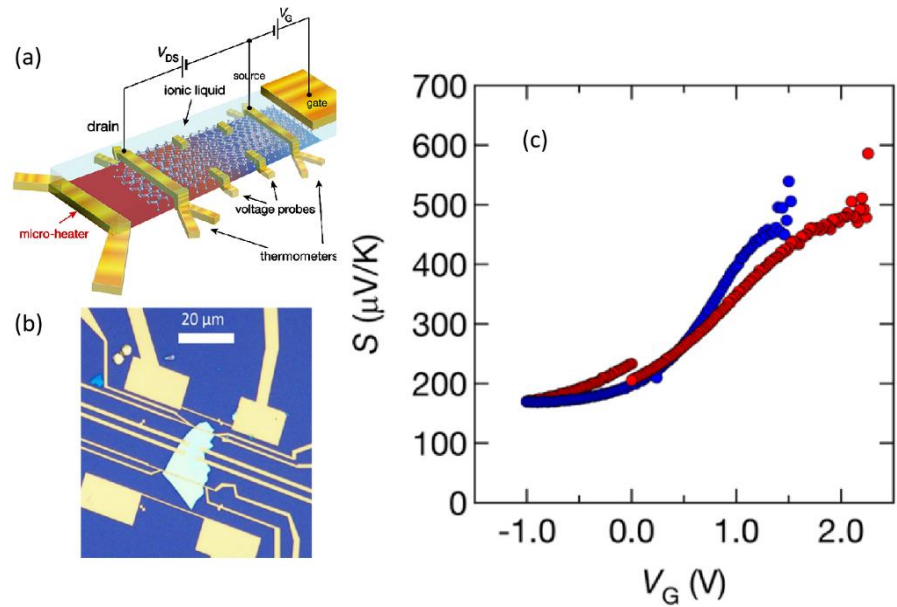

**Fig. 7** (a) is the schematic diagram of modifying the Seebeck coefficient of black phosphorus by ion gating. (b) The actual optical image corresponding to the device in (a). (c) Experimental result of the dependence of Seebeck coefficient ($S$) on gate voltage ($V_G$). Reproduced from Ref. [68].

### 3.2.2 Directly mapping Seebeck coefficient

Apart from the above theoretical calculation and the experimental electrical measurement method developed by Kim *et al*. to obtain the global Seebeck coefficient, there is existing apparatus that can directly map the Seebeck coefficient distribution of sample surface, which provides indepth understanding of the origin of variation of local Seebeck coefficient. The first one that is capable of imaging the Seebeck coefficient distribution on the sample surface was proposed by Liang *et al*. in 2004, and the operating principle is mainly based on scanning tunneling microscope (STM) and termed as scanning thermoelectric microscope (SThEM) (Figure 8(a)) [71]. When the tip is in contact with the sample, a local thermovoltage is generated due to the temperature difference, and electrical contact between the tip and the sample is required, then the Seebeck coefficient distribution can be derived from the measured thermovoltage. Using this technique, the Seebeck coefficient distribution profile on a gallium arsenide (GaAs) p-n junction sample was obtained. Later, Cho *et al.* employed this apparatus and Park *et al.* applied simply scanning tunneling microscopy (STM) to probe the thermoelectric properties of graphene grown on SiC. Both research results show that the number of layers and the structural disorders formed during the growth process has a great influence on the thermoelectric properties of epitaxially grown graphene (Figure 8(b)) [72-74]. Lately, Harzheim *et al.* developed an instrument called scanning thermal gate microscopy (STGM) (Figure 8(c)), they utilized this instrument to comprehensively and systematically study the effects of structural defects, metallic contacts, thickness and electric field-effect regulation on the Seebeck coefficient of mechanically exfoliated graphene, and the instrument does not require electrical contact between the tip and the sample [75]. Although the above-mentioned three test methods of tip scanning Seebeck coefficient distribution are similar to the aforementioned PET, they have the following two advantages: the first is atomic resolution and the second is that it will not be confused with other optical effects [47]. It can be seen that thermovoltage scanning instruments are a very favorable means, which can obtain some electronic structural features that cannot be acquired through general electrical measurement methods, thus facilitating us to understand the underlying mechanism of thermoelectric performance and optimize and improve its properties.

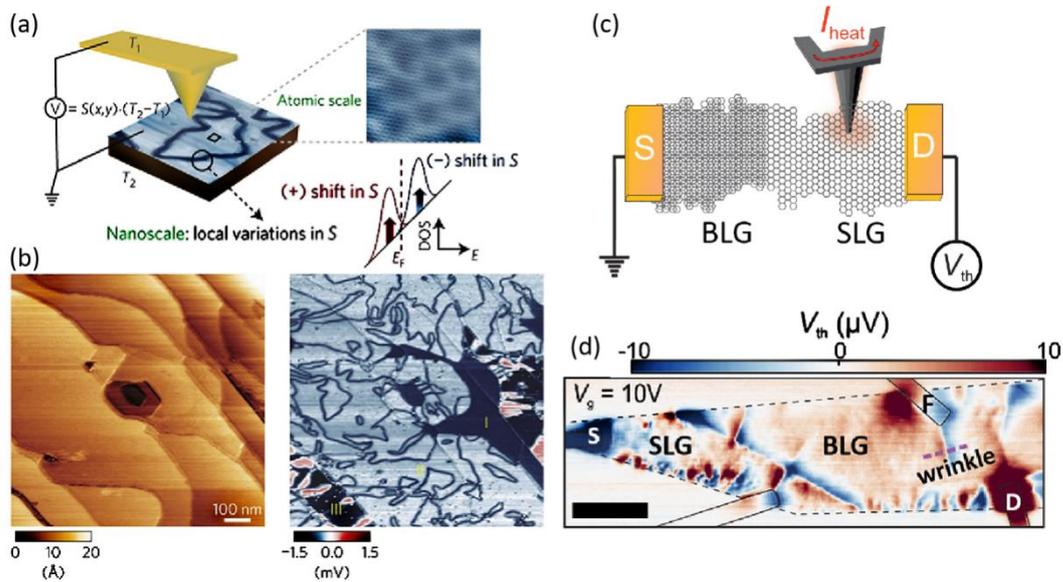

**Fig. 8** (a) The experimental setup of scanning thermoelectric microscope (SThEM). (b) Topography and thermovoltage maps of epitaxial graphene with different thicknesses. (c) Schematic diagram of scanning thermal gate microscopy (STGM) and thermovoltage map (d) of a single-layer (SLG) and bilayer (BLG) graphene flake under a fixed gate voltage. Reproduced from Refs. [73, 75].

### 3.2.3 Thermoelectric effects in nanoscale homojunctions

In the following part, we first introduce two examples of imaging thermoelectric effects in graphene with SThM, where the traditional heterojunction structures are not involved. The first study was carried out by Harzheim *et al.* [76], using the method proposed by Menges *et al.*. The experimental results are illustrated in Figure 9(b). A striking phenomenon observed near the 100 nm wide graphene bow-tie nanoconstriction, that is, the Peltier heating and cooling signals are locally distributed on both sides of the nanoconstriction and the resulting temperature difference can reach about 2K, which was not unexpected. And the Joule heating signal follows the Joule-Lenz law and concentrates on the nanoconstriction where the current density is the highest. The research also detected the thermovoltage distribution by actively heating the tip and the result shows that the Seebeck coefficient has changed around the nanoconstriction, which is consistent with the observed Peltier signal. They interpreted this extraordinary phenomenon as the modification of Seebeck coefficient by a classical size effect, namely, the electron mean free path (EMFP) decreases with the decrease of channel width and the narrowest channel region (~100 nm) has a minimum Seebeck coefficient. Due to the symmetry of the device structure, the dip of the Seebeck coefficient in the constricted region eventually leads to the remarkable Peltier image. It is note that in this research, although the Au contact pad is typically very large, the current density at the heterointerface of graphene/Au is very small (much smaller than the nanoconstriction region). Hence no substantial thermoelectric effect is discernible at the interface of electrodes and the thermoelectric signal from the nanoconstricted homojunction region dominates.

Another example is about the effect of the wrinkles formed during the growth of germanium-based graphene on the microscopic thermoelectricity, studied in our research group jointly [77]. The experimental setup is illustrated in Figure 9(a). In this work, we first investigated the thermoelectric effects of a 2 μm wide graphene channel with randomly distributed wrinkles as a result of the mismatch of thermal expansion of the graphene and germanium substrates. Peltier heating and cooling occurred locally on both sides of the wrinkles and Joule heating happened to appear right at the wrinkles. The experimental results are also confirmed by thermovoltage imaging obtained with STGM [75]. Inspired by the size effect proposed by Harzheim *et al.*, we combined nano-bubble engineering with wrinkles to increase the current density through the wrinkles for the purpose of increasing the intensity of the Peltier signal. Compared with the traditional electron beam exposure technology, the nano-bubble engineering is more convenient, flexible and less difficult to operate. When the AFM tip is applied a negative voltage relative to the sample, the Ge-H at the interface between the graphene and germanium will turn into hydrogen molecules, thereby forming bubbles. The corresponding topography and thermoelectric images are displayed in Figure 9(c). Benefiting from this special geometry structure, the measured thermoelectric signal has been greatly enhanced. Besides, we have also observed the decoupled Peltier and Joule signal, that is, the transition of Peltier cooling and heating is not exactly at the same position as the maximum value of Joule heating. The former tends to be located at the positions of the wrinkles, while the latter tends to be concentrated in the narrowest part of the channel where wrinkles sit nearby. We attribute the unique thermoelectric properties of graphene wrinkles to the hole transport through the van der Waals barrier of the graphene wrinkles which filters the energy-dispersive charge carriers in the transport.

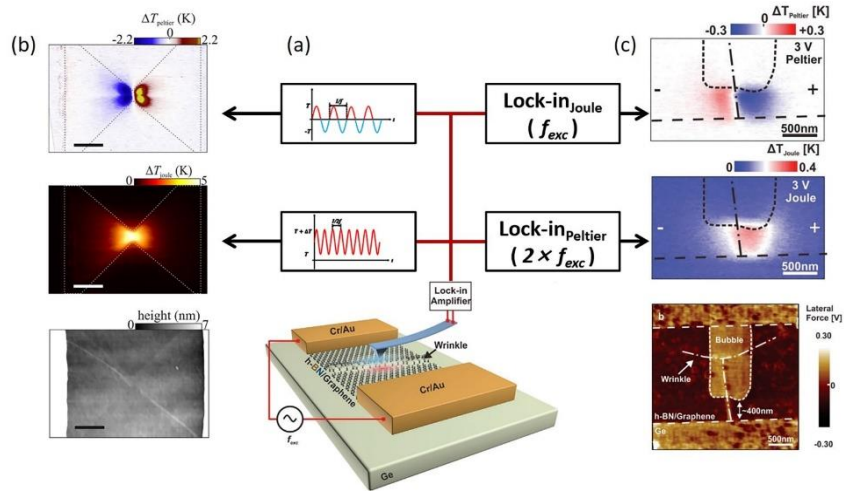

**Fig. 9** (a) Schematic diagram of the experiment using SThM to measure the thermoelectric effects in a graphene device with a nanoconstriction structure (b) and a wrinkle graphene device made from bubble-engineering technology (c). (b) and (c) from top to bottom corresponding to the measured Peltier, Joule and topography maps, respectively. Reproduced from Refs. [76, 77].

Another interesting thermoelectric system is the phase-changing nanowires with non-uniform configuration. Vanadium dioxide ($VO_2$) is a strongly correlated material and it can be reversibly transformed between the metal and the insulator phase by external heating, stress, self-heating driven by external current [78-80]. A plenty of studies have shown that the phase transition process of $VO_2$ is gradual and uneven, so inevitably forming many domain walls [81-83]. Favaloro et al. first tried to use thermoreflectance imaging microscopy with 200-300 nm spatial and 10 mK temperature resolution to study the temperature profile across the interface between metal and insulator in a coexisting phase regime of $VO_2$ (Figure 10(a)) [84]. The measured temperature magnitudes and related phase information of the Peltier and Joule signals are illustrated in Figure 10(b) and 10(c), respectively. It is apparent that both the signals are located around the interface between metal-insulator domains with the domain walls acting as an energy barrier in highly resistive condition. From the experimental results, the Peltier phase image shows that there is a 180° phase difference between the two phase transition junctions which is remarkably different from the Joule phase diagram (the lower panels of Figure 10(b) and 10(c)). Lately, Könemann et al. made use of SThM to investigate the thermoelectric effects at crystal phase junctions, wurtzite (WZ) / zinc-blende (ZB) / wurtzite (WZ), in a single InAs nanowire with average diameter about 60nm and the WZ section length around 82 nm (Figure 10(d)) [85]. Results are displayed in Figure 10(e) and 10(f). Peltier heating and cooling and Joule heating maximum are distributed around the interface of WZ/ZB and the line profiles (Figure 10(g)) is more obvious. The reason for this phenomenon is that InAs nanowire with different crystal phases has different energy bands, therefore, leading to distinctive Seebeck coefficients [86]. In the above two researches showing the coexistence of multi-phases in nanowires, different measurement techniques, i.e., the thermoreflectance imaging microscopy and SThM, were used respectively. From their results, two methods show consistent data despite that SThM exhibits a superior resolution than thermoreflectance imaging microscopy.

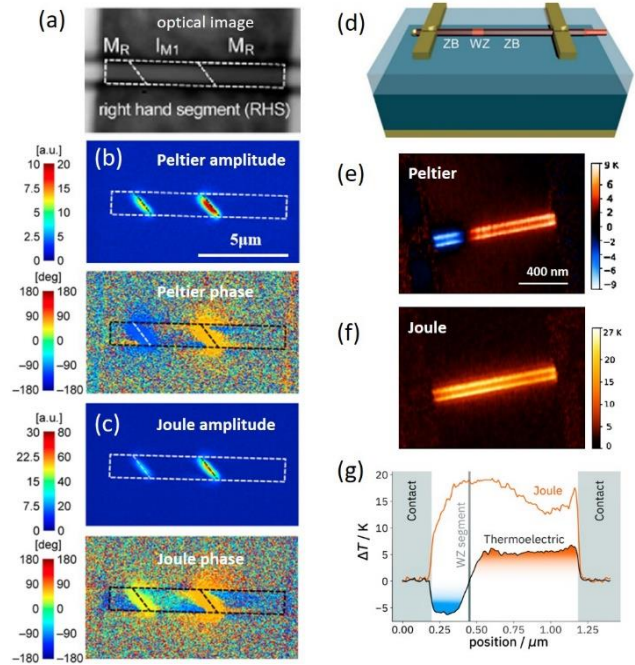

**Fig. 10** (a) The optical image of a multiphase mixed vanadium dioxide nanowire. (b) The Peltier amplitude (upper panel) and phase distribution (lower panel), (c) Joule amplitude (upper panel) and corresponding phase signal (lower panel) at the phase interface of a multiphase mixed $VO_2$ nanowire under 8 μA current excitation measured by the thermoreflectance thermal imaging microscopy. (d) Schematic diagram of a single InAs nanowire with mixed phase structure. The Peltier (e) and Joule (f) images of a single InAs nanowire under 1V voltage amplitude modulation measured by SThM and the line profiles are exhibited in (g). Reproduced from Refs. [84, 85, 86].

## 3.3 Spin thermoelectric effects

Recently, a novel research field called spin caloritronics has caught great attention, holding great promise to increase the efficiency and the versatility of the thermoelectric devices, which combines conventional heat transfer with the spintronics [87]. In 2008 and 2010, Uchida *et al.* successively observed the spin Seebeck effect in magnetic metal and magnetic insulator and in experiments, the inverse spin Hall effect (ISHE) was used to successfully convert the spin current generated by temperature gradient in magnetic metal or magnetic insulator into a measurable voltage, thus ingeniously verifying the existence of spin Seebeck effect (SSE) [88, 89]. After that, in 2012, Flipse *et al.* used the thermocouple measurement method to study the spin-dependent Peltier effect in a sandwich structure composed of Py/Cu/Py [90]. Four years later, Daimon *et al.* first explored the spatial temperature distribution in the magnetic junction systems composed of a layer of paramagnetic metal (PM) Pt and a layer of ferrimagnetic insulator (FI) YIG by lock-in thermography (LIT) [91]. The device structure and the experimental set-up are shown in Figure 11(a), a similar AC test method mentioned above is adopted, except that the sine wave is replaced with a square wave, and the measured Peltier signals are divided into amplitude and phase images. When a current $J_C$ is input into a PM with strong spin-orbit coupling, the spin Hall effect (SHE) will generate a spin current and then accumulate at the PM/FI interface, and the accumulated spin current will combine with the magnetic moment in the FI through the interfacial exchange interaction, thus the spin angular momentum and energy transfer can be carried out between the electrons in PM and the magnons in FI (Figure 11(b)). The heat flows through the interface has the following form $J_q \propto (\sigma \cdot M)n \propto J_c \times M$. As shown in the Figure 11(c), the spin Peltier signal appears in the L and R area, and the corresponding phase has a 180° difference, and there is no signal in the C area. It is worth pointing out that the temperature distribution caused by the spin-Peltier effect is mainly localized at the interface, with only a small thermal diffusion, and the temperature can be flexibly and accurately controlled by an external magnetic field and the excitation current, which therefore has great potential in thermoelectric applications.

Soon after, the same research group observed anisotropic magneto-Peltier effect (AMPE) in a single ferromagnet material Ni with U-shaped structure without traditional heterojunction structure by means of LIT [92]. They take advantages of the anisotropic magento effects which refer that the Peltier coefficient can be tuned through modulating the angle between the direction of the applied charge current and the magnetization vector of the ferromagnet due to spin-orbit interaction, thus forming a non-uniform magnetizaiton configuration (Figure 11(e)). Refer to the above-mentioned nanowires with inhomogeneous phase composition, there are similarities and the same effect. The experimental configuration and some representative results are shown in Figure 11(d). As expected, since the current flow direction and the magnetization vector in areas $B_{L/R}$ and $B_C$ in the figure are perpendicular and parallel, respectively, so the AMPE should appear at the junction, that is, area $C_{L/R}$. The phases of area $C_L$ and $C_R$ are 180° difference which reveals that heat is absorbed or released at $C_L$ or $C_R$. Additionally, in the original article, they also systematically studied the influence of the magnitude of the applied current, the magnitude of the magnetic field and the angle between the magnetic field and the current on the AMPE. Besides, they also put forward a novel idea that the magnetic field control degree of freedom can freely form a non-uniform magnetization composition in a ferromagnetic material to achieve the AMPE, which will facilitate its application in heat management. With the in-depth study of spin caloritronics and the trend toward smaller-scale development, it is believed that SThM may find potential applications in the research field of spin caloritronics.

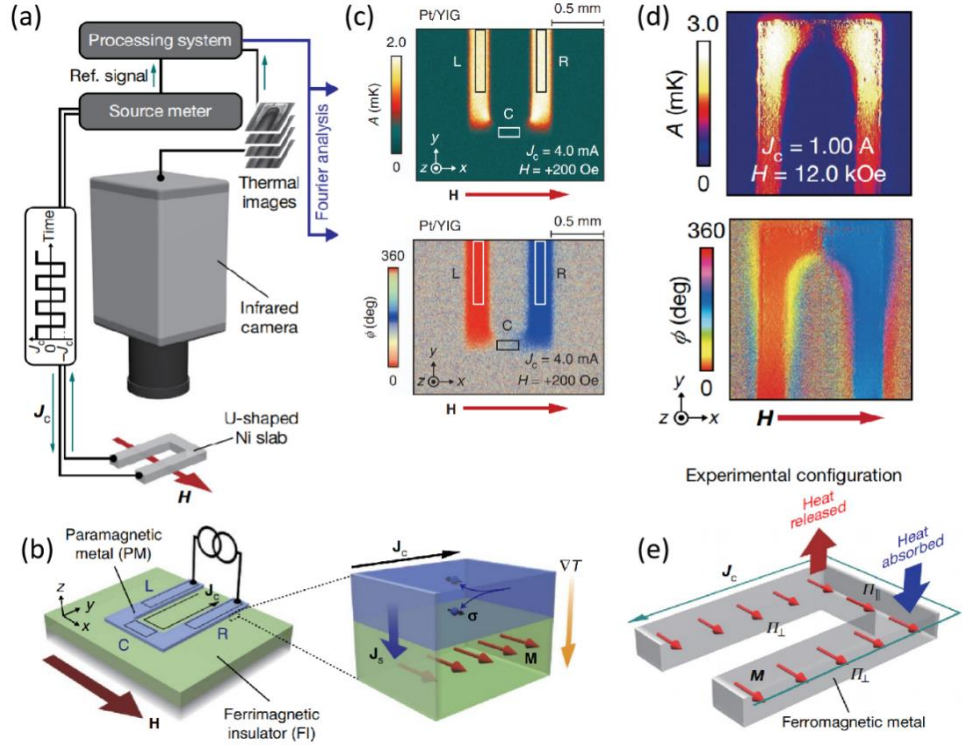

**Fig. 11** (a) Schematic diagram of the experimental setup for measuring the spin Peltier effect with an infrared emission microscopy. (b) Illustration of the device composition and the principle of the Peltier effect generated at the interface of ferromagnetic insulator (FI) and paramagnetic metal (PM). The measured results in PM/FI interface and a single ferromagnetic material Ni are displayed in (c) and (d). (e) The experimental configuration for producing Peltier effect in a single ferromagnetic material. Reproduced from Refs. [91, 92].

## 4. Probing Non-equilibrium transporting charges with SNoiM

The above progress in the microscopic thermoelectric studies greatly improves our knowledge of the underlying mechanisms. On the other hand, all those experiments are restricted to or at least dominated by the detection of lattice heat. Thermoelectric effects, however, arises intrinsically by the non-equilibrium charge transport, which remains extremely challenging to be directly visualized. In this section, we introduce our recently developed Scanning Noise Microscope (SNoiM) for direct probing of non-equilibrium transporting hot charges (instead of lattice heat) in semiconductor nanodevices and hopefully in thermoelectric devices in near future [93].

SNoiM is a kind of nearfield scanning optical microscope (NSOM), which can collect terahertz electromagnetic fluctuations (noises)generated by hot electrons in sample with an ultrahigh sensitive detector, and super resolution is achieved by a very sharp metal tip which can scatter the fluctuating electromagnetic evanescent field on the sample surface. Therefore, SNoiM can be regarded as a near-field version of the microscopic infrared thermometer, as schematically shown in Figure 12(a). Compared with SThM in contact mode, SNoiM has a promising spectroscopic freedom despite that currently it works only at single terahertz frequency (~21THz). The system works in a non-contact mode or tapping mode to subtract the influence from strong black-body background in this terahertz range. To avoid the disturbance from lattice environment, it is crucial that, the detector collects only a narrow band of electromagnetic waves which is well away from any phonon resonances of the hosting lattice, such that the total electromagnetic local density of states (EM-LDOS) is dominant by the contribution from electrons. As a result, SNoiM probes the electron dynamics.

In a nano channel fabricated on GaAs semiconductor device, SNoiM provides for the first time the real space image of hot electron temperature distribution and surprisingly non-local energy dissipation of hot electrons is observed despite that the GaAs devices are located at ambient temperature (~300K) [94]. As shown in Figure 12(b), the electron hot spot reaches as high as ~2000K which is much higher than lattice and hence very far away from thermal equilibrium with lattice. From these results, it is expected that SNoiM can be hopefully employed in exploring the rich dynamics of nonequilibrium phenomena in a broad range of material and device systems including the thermoelectrics.

It is worth pointing out that SNoiM and SThM provide complementary information about the electron and lattice which is both important in thermoelectric applications. A preliminary simulation in a GaAs U-shaped channel shows that the electron and lattice temperature distributions can exhibit remarkably different patterns as depicted in Figure 12(c) and (d). A direct comparison in experiments is highly anticipated to confirm this difference and is under the way. Hopefully, SNoiM together with SThM will be able to provide more complete and comprehensive understanding to the microscopic mechanisms in thermoelectrics.

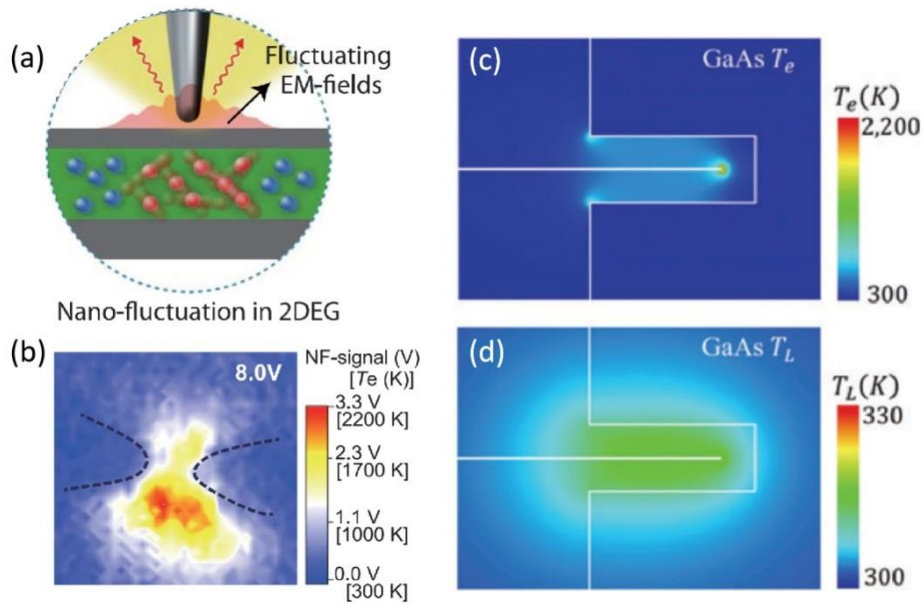

**Fig. 12** (a) Schematic diagram of the nano-fluctuation in GaAs 2DEG. (b) Two-dimensional excess noise distribution image of a GaAs 2DEG device measured by scanning noise microscopy (SNioM). (c) and (d) are the simulated temperature distribution of electrons and lattice of U-shaped n-GaAs device, respectively. Reproduced from Refs. [93, 94].

Table 1 The including material system discussed in this review.

| Classification | Composition | Principle | Performance | Measurement method | Citation |
|---|---|---|---|---|---|
| Heterojunction | Graphene/Au | Different Seebeck coefficients | $|\Delta T_{Peltier}| \approx 1K$ @ $I \approx 0.6mA$ / $|\Delta T_{Peltier}| \approx 15mK$ @ $I \approx 20\mu A$ | SJEM/ Electrical | Ref. [42, 44] |
| | GST/Au | Different Seebeck coefficients | $|\Delta T_{Peltier}| \approx 3K$ | SJEM | Ref. [43] |
| | InAs/Au | Different Seebeck coefficients | $|\Delta T_{Peltier}| \approx 1.5K$ | SThM | Ref. [50] |
| Single material /Homojunction | Metal thin film strip | Size effect of Seebeck coefficient | NA | Electrical | Ref. [58, 59] |
| | Graphene nanoconstriction | Size effect of Seebeck coefficient | $|\Delta T_{Peltier}| \approx 2K$ @ $I \approx 90\mu A$ | SThM | Ref. [76] |
| | Graphene wrinkle | Energy filtering of the tunnel barrier | $|\Delta T_{Peltier}| \approx 0.226K$ @ $I \approx 107\mu A$ | SThM | Ref. [77] |
| | $VO_2$ nanowire | Different Seebeck coefficients due to Multiphase | $\frac{\Delta T_{Joule}}{|\Delta T_{Peltier}|} \approx 1.75$ @ $I \approx 10\mu A$ | Thermoreflectance | Ref. [84] |
| | InAs nanowire | Different Seebeck coefficients due to Multiphase | $|\Delta T_{Peltier}| \approx 5K$ | SThM | Ref. [85] |

## 5. Conclusion and Outlook

In this short review, we briefly recall the recent progress of microscopic thermoelectric research carried out on different material systems including the traditional heterojunction structures (e.g., Graphene/Au, GST/Au etc.), single-material homojunctions (e.g., bowtie-shaped graphene, multi-phase nanowires), and also spin Peltier in magnetic materials. Table 1 summarizes the main material systems reviewed in this article, which are classified as heterojunction and homojunction structures and compared in terms of principle and performance for thermoelectric effects. As the potential and promising candidates for thermoelectricity applications in solid-state electronics, these low dimensional or nano-materials are supposed to be used in integrated industry and their physical properties require to be intensively studied. Their engineered Seebeck coefficient as well as thermal conductivity will lead to a great increase in the thermoelectric and/or thermal power figure of merit and hence facilitate actual applications. To achieve this aim, a comprehensive understanding of microscopic thermoelectric mechanisms is essential and the rapid development of the experimental techniques (e.g, SThM and SNoiM) brings new opportunities to this frontier research field.

## Acknowledgement

We are grateful to Profs. Wei Lu, and Zengfeng Di from Chinese Academy of Sciences, and Prof. S. Komiyama from University of Tokyo, Drs. Qianchun Weng, Le Yang and Yuexin Zou, Xudong Hu for very fruitful collaborations. We acknowledge funding support from Shanghai Science and Technology Committee under grant Nos. 20JC1414700, 18JC1420402, 18JC1410300, National Natural Science Foundation of China (NSFC) under grant Nos. 11991060/11674070/11634012, and National Key Research Program of China under grant No. 2016YFA0302000.